\documentclass[a4paper]{article}

\usepackage{INTERSPEECH2022}
\usepackage{tabularx}
\usepackage{hyperref}
\usepackage{float}
\usepackage{placeins}

\title{tPLCnet: Real-time Deep Packet Loss Concealment in the Time Domain Using a Short Temporal Context}
\name{Nils L. Westhausen$^{1}$ and Bernd T. Meyer$^{1}$}

\address{
  $^1$Communication Acoustics \& Cluster of Excellence Hearing4all\\
  Carl von Ossietzky University, Oldenburg, Germany}
\email{nils.westhausen@uol.de, bernd.meyer@uol.de}

\begin{document}

\maketitle
\begin{abstract}
This paper introduces a real-time time-domain packet loss concealment (PLC) neural-network (tPLCnet). It 
efficiently predicts lost frames from a short context buffer in a sequence-to-one (seq2one) fashion. 
Because of its seq2one structure, a continuous inference of the model is not required since it can be triggered when packet loss is actually detected.
It is trained on 64\,h of open-source speech data and packet-loss traces of real calls provided by the Audio PLC Challenge. 
The model with the lowest complexity described in this paper reaches a robust PLC performance and consistent improvements over the zero-filling baseline for all metrics. 
A configuration with higher complexity is submitted to the PLC Challenge and
shows a performance increase of 1.07 compared to the zero-filling baseline in terms of PLC-MOS on the blind test set and reaches a competitive 3rd place in the challenge ranking.
\end{abstract}
\noindent\textbf{Index Terms}: real-time, packet loss concealment,  RNN

\section{Introduction}
Several crucial processing steps for virtual communication have immensely profited from deep learning, e.g., speech enhancement (SE) \cite{perceptnet_dns, dtln_dns} and acoustic echo cancellation (AEC) \cite{perceptnet_aec, dtln_aec}. 
The underlying models can be created with deep-learning tools alone or based on a combination with ''classic`` signal-processing approaches such as adaptive filters and with deep-learning models \cite{perceptnet_aec, aec_braunschweig}. 
Another important challenge is to improve communication systems with effective packet loss concealment (PLC) using deep learning.
The task of PLC is to conceal or predict lost packets during packet-based transmissions. A well-performing PLC system is especially important for poor network connections, since more packets are lost or overly delayed and therefore discarded. A robust PLC system could also decrease listening fatigue and effort in situations with poor audio quality and therefore result in overall improved communication channels. PLC systems should work efficiently so they can be used for many device classes without fully utilizing the processing power and sacrificing battery run-time on mobile devices. 

Packet loss concealment is a sequence problem since missing frames from an audio sequence must be predicted from past information. 
An early deep approach \cite{dnn_packet_loss} utilized feed-forward neural networks for predicting the magnitude and phase from a number of past frames using mean-squared-error based loss in time-frequency domain. 
Another approach is a causal sequence to sequence model such as RNNs as used for SE \cite{nsnet2, dtln_dns, cruse}. In related research, RNN-based models have been applied to PLC \cite{mohamed2020concealnet, plc_lstm, facebook_plc}. In \cite{plc_lstm} a recurrent model with online adaptation is proposed by exploiting the recent past of the signal. \cite{mohamed2020concealnet} used an LSTM-based PLC model to increase the performance of speech emotion recognition trained with a concordance correlation coefficient as cost function. A convolutional recurrent structure for time-domain PLC was introduced in \cite{facebook_plc} trained with a mean absolute error cost in time domain.
Alternative deep approaches utilizing generative adversarial networks (GAN) \cite{plc_gan_1, inpainting_gan, plc_gan_2}, which were shown in the past to be able to predict speech frames. 

A comparison between these studies is difficult because of the lack of a common training and test set, and therefore Microsoft proposed a deep PLC Challenge for Interspeech 2022 \cite{diener2022interspeech}. 
The authors provided a large amount of speech data, traces of lost frames from real calls and a 
deep model for predicting mean opinion scores (MOS), indicating PLC performance of enhanced audio. The challenge evaluations will be performed as in previous challenges organized by Microsoft \cite{dns_challenge, AEC_challenge_ver1, dns_2021_interspeech} 
with a crowd-sourcing approach on the Amazon Mechanical Turk.

For the first time, to the best of our knowledge, we propose in the context of the PLC Challenge a time-domain sequence-to-one (seq2one) RNN model for PLC trained with a combined magnitude and complex mean absolute error (MAE) loss in time-frequency domain. In contrast to previous RNN approaches \cite{mohamed2020concealnet, plc_lstm, facebook_plc}, the model predicts only lost frames from a short context buffer similar as in \cite{dnn_packet_loss} in a non-continuous way instead of running inference for each incoming audio frame to update internal states. This seq2one approach can decrease, depending on the complexity of the model, the computational load of the host system
and probably increase the battery run-time of a mobile device.
\section{Methods}
\subsection{Deep PLC Strategy}
\begin{figure}[t]
  \centering
  \includegraphics[width=167pt]{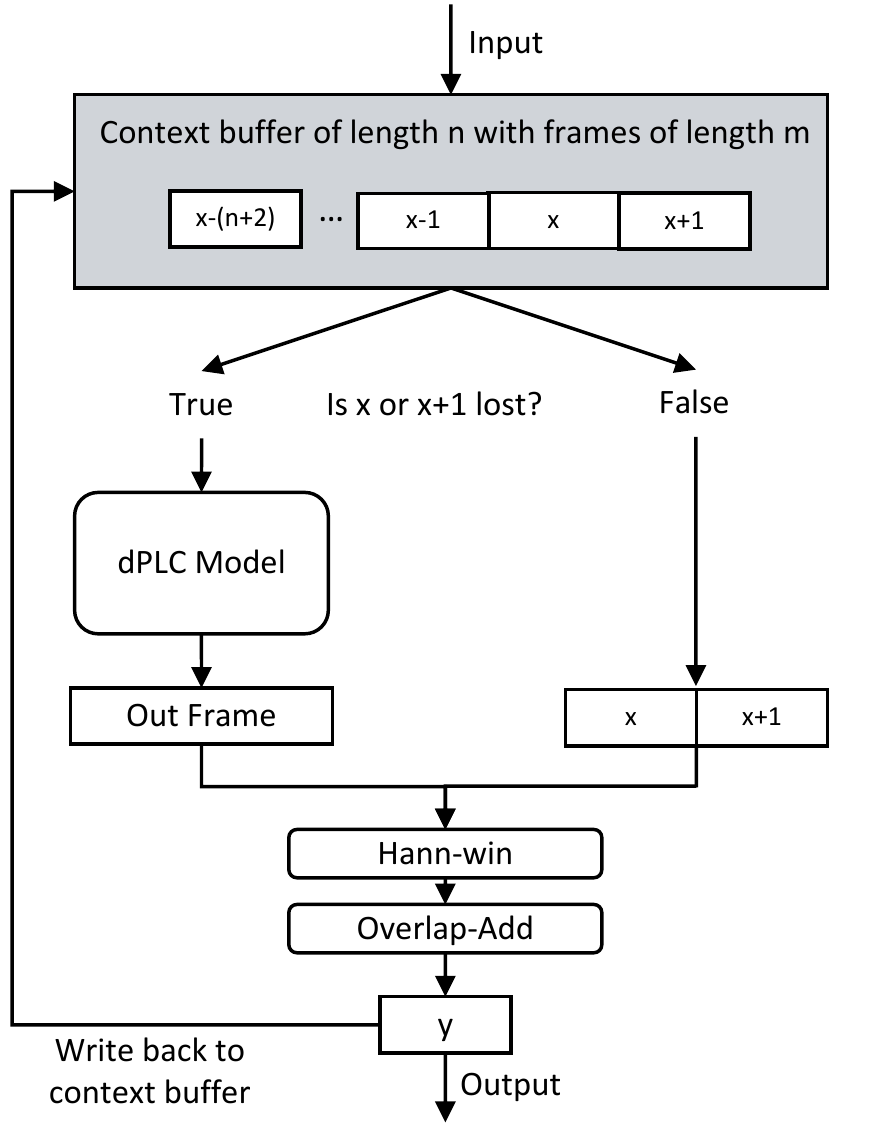}
  \caption{Illustration of the proposed PLC system. Incoming audio is written to a context buffer. When no frame is lost, the current frame and the look-ahead frame are concatenated and hann-windowed. If the the frame on position $x$ or $x+1$ is lost, the dPLC predicts an audio frame which is also hann-windowed. The resulting windowed frames are processed by an overlap-add procedure returning the current frame $y$ time aligned to input frame $x$.}
\label{fig:plc-system}
\end{figure}

The proposed PLC seq2one system is based on a context buffer of length $n$ that contains non-overlapping frames of length $m$. 
For $m$ 10\,ms (160 samples at 16\,kHz sampling rate) are chosen. It is therefore possible to use one look-ahead frame and still comply with the PLC-Challenge rules, which specify a maximum delay of 20\,ms \cite{diener2022interspeech}. The use of look-ahead frames has been shown to be beneficial in \cite{facebook_plc} to connect predicted frames better to available audio. When no packet loss is detected, the current and the look-ahead frames are concatenated to a frame with 320 samples and subsequently hann windowed. 
When a packet loss is detected, the dPLC model predicts a 320 sample long frame that is windowed as well. 
With an overlap-add procedure the current 10\,ms output frame is predicted which is time-aligned to the current input frame. 
The current output frame is written back to the context buffer.
This procedure, especially the windowing and overlap-add, was chosen to reduce artifacts on the edges of lost packets and connect the predicted content better to the available frames and so pursues a similar goal as the cross-fading shown in \cite{dnn_packet_loss}. The proposed PLC system is also illustrated in \autoref{fig:plc-system}.
\subsection{Model Architectures}
The model is build as seq2one model predicting one time domain frame from a context buffer. The general architecture is inspired by the NSnet2 architecture \cite{nsnet2} and an architecture proposed in \cite{facebook_plc}. It was adapted to seq2one task and working completely in the time domain.

First, the frames of the context buffer are processed by a fully-connected (FC) layer with relu activation that maps the time-domain frames to a non negative feature space, as has been proposed with the TasNet approach \cite{tasnet_orig,conv_tasnet, dua-path}. 
Second, all frames are processed by another FC layer with leaky-relu activation. The resulting intermediate representation sequence is processed by two consecutive Conv1D layers with leaky relu-activation and kernel size 4 and 2, to extract local patterns from the sequence. The sequence is padded so that the resulting sequence has still the same length. 
The sequence is subsequently processed by two bidirectional Gated Recurrent Unit (GRU) \cite{chung2014empirical} layers, which enables the use of information about the whole sequence. The final hidden state of the second BGRU is forwarded to two FC mapping layers with leaky-relu activation. A last FC layer without activation predicts the time-domain frame. 
To construct a feed-forward baseline (FF-Baseline) the Conv1Ds and BGRUs layers are replaced by a flatten operation to remove the sequence dimension and 3 FC layers with 512 units and leaky-relu activation.

In total, three different size configurations are tested for the proposed tPLCnet to explore the scaling properties of the model (\emph{Small, Medium and Large}). The number of units or filters are listed in \autoref{tab:units}. Additionally, \autoref{tab:results} shows the number of Multiply Accumulates (MACs) and execution time for predicting one 10\,ms frame on an Intel i5-7267U dual-core processor clocked at 3.1\,GHz.
\begin{table}[th]
  \caption{Number of units/filters of each layer for the configurations \emph{Small, Medium and Large} of the proposed deep PLC model.}
  \label{tab:units}
  \centering
  \begin{tabular}{ l c c c}
    \toprule
    \textbf{Layer} & \textbf{Small} & \textbf{Medium} & \textbf{Large} \\
    \midrule
     FC encoding &   \multicolumn{3}{c}{512} \\
     \midrule
     FC embedding  &   128   &     256  & 512 \\
      \midrule
     Conv1x4 &   128    &     256  & 512 \\
      \midrule
     Conv1x2 &   128  &     256  & 512 \\
      \midrule
     BGRU 1 &   64  &     128  & 256 \\
      \midrule
     BGRU 2 &   64  &     128  & 256 \\
      \midrule
     FC mapping 1 &    \multicolumn{3}{c}{512} \\
      \midrule
     FC mapping 2 &    \multicolumn{3}{c}{512} \\
      \midrule
     FC decoding &    \multicolumn{3}{c}{320} \\
    \bottomrule
  \end{tabular}
\end{table}
\subsection{Loss Functions}
Recent studies have shown that the loss function is crucial to perceived audio quality in speech enhancement \cite{loss_functions_1}, but at the same time, relatively simple loss functions are surpassing more complex ones \cite{loss_functions_2}. 
To quantify the effect loss functions for PLC, our study compared three loss functions, which are all applied to the whole enhanced utterance.

\cite{facebook_plc} proposed to use a simple mean absolute error (MAE) in time domain which shown promising results. 
\begin{equation}
    \mathrm{MAE_{time}} = |\hat{x} - x|
\end{equation}
where the loss calculated as the mean absolute difference of the clean audio $x$ and the predicted audio $\hat{x}$.
Due to its definition in time domain, MAE is phase sensitive and is - in comparison to the mean squared error - less sensitive to outliers.  

In \cite{loss_functions_1}, it was shown that the MAE in time-frequency domain (TF-domain) produces competitive results for speech enhancement. 
MAE has also been explored for speech synthesis  \cite{wang17n_interspeech}, which in some aspects is related to PLC.
To calculate the MAE in TF-domain, the clean and predicted utterances are subject to an STFT which results in the complex TF representations $X(t,f)$ and $\hat{X}(t,f)$. 
The magnitude MAE loss is given by
\begin{equation}
    \mathrm{MAE_{mag}} = ||\hat{X}| - |X||.
\end{equation}
Results of related studies indicate that integrating a phase component in the loss function can be beneficial \cite{psa_1, loss_functions_1}, which can be implemented by using a complex distance. The combined TF-domain MAE loss is given by
\begin{equation}
    \mathrm{MAE_{comb}} = (1-\alpha) ||\hat{X}| - |X|| + \alpha |\hat{X} - X|,
\end{equation}
where $\alpha$ is a weighting factor for the amount of phase sensitivity. In this study, $\alpha = 0.1$ is chosen, which was found to be a good trade-off between some phase sensitivity and a good magnitude reconstruction in preliminary experiments. 
\begin{table*}[!th]
\caption{Results on the known test-set in terms of PESQ [MOS], PLC-MOS [MOS] and DNS-MOS P.835. For PESQ and PLC-MOS the data is also shown for the three loss subsets low (up to 120\,ms), med. (120 to 320\,ms) and high (320 to 1000\,ms). For DNS-MOS all three metrics are shown. Additionally the number Multiply Accumulates (MACs) and the execution time [ms] for one 10\,ms frame are shown.}
  \centering
  \begin{tabular}{@{}l l r r r r r r r r r r r r r r@{}}
  \toprule
  & & \multicolumn{1}{c}{\textbf{exec.}} &  \multicolumn{4}{c}{\textbf{PESQ}} & & \multicolumn{4}{c}{\textbf{PLC-MOS}} &  & \multicolumn{3}{c}{\textbf{DNS-MOS}}
        \\
    \midrule
    \textbf{Method}  & \textbf{\# MACs}  & \textbf{Time}  & low &  med.   & high & ovr. &
    & low &  med.   & high & ovr. & & \textbf{nSIG}   & \textbf{nBAK}   & \textbf{nOVR}
     \\
     \midrule
     Baseline & & & 2.59 & 1.75 & 1.73 & 2.19 & & 3.15 & 2.46 & 2.77 & 2.87 & & 3.47 & 3.61 & 2.97 \\
     \midrule
     FF baseline & 2.49\,M & 0.47  & 3.24 & 2.13 & 1.91 & 2.68 & & 4.05 & 3.52 & 3.63 & 3.81 & & 4.15 & 4.23 & 3.75 \\
     $\mathrm{tPLCnet_{S}}$ & 2.96\,M & 0.57 & 3.28 & 2.14 & 1.97 & 2.71 & & 4.09 & 3.57 & 3.72 & 3.86 & & 4.14 & 4.23 & 3.74 \\
     $\mathrm{tPLCnet_{M}}$ & 7.85\,M & 1.41 &   3.31 & 2.15 & 1.98 & 2.74 & & 4.12 & 3.61 & 3.79 & 3.90 & & 4.17 & 4.25 & 3.78 \\
     $\mathrm{tPLCnet_{L}}$ & 26.18\,M & 3.95 &  \textbf{3.36} & \textbf{2.18} & \textbf{1.99} & \textbf{2.77} & & \textbf{4.16} & \textbf{3.66} & \textbf{3.82} & \textbf{3.95} & & \textbf{4.18} & \textbf{4.27} & \textbf{3.79} \\
    
     \bottomrule
  \end{tabular}
  \label{tab:results}
\end{table*}
\subsection{Datasets and Augmentation}
The challenge organizers are providing a training set, a known test set and a blind test set. The training set contains around 64\,h of clean speech and packet loss traces from real calls. 
The traces are vectors of ones and zeros indicating for each 20\,ms frame/packet of an audio file if it is lost (one) or not (zero). 
The speech is sampled from public available podcast audio. Each utterance is around 10\,s long. 
During training, the speech files and PLC traces are cut to 8\,s with a random start and end. The order of speech files and traces are shuffled for each epoch. The traces are time-reversed with 50\,\% probability to further increase the variance and increase robustness. To create the degraded audio the lost segments are set to zero according to the corresponding trace. The level of the speech is sampled from a normal distribution with a mean of -26\,dB and a standard deviation 10\,dB.  

The known test set contains around 2.7\,h of clean speech, degraded speech and PLC traces. As in the training set, each file is around 10\,s long. The data is divided in three subsets by the organizers depending on the maximum burst loss length of the traces. The categories are: low loss up to 120\,ms (52\,\% of the data), medium loss from 120 to 320\,ms (32\,\% of the data) and high loss from 320 to 1000\,ms (16\,\% of the data). 

The blind test set has the same properties, but does not contain the clean speech files. Further details on the data can be found in the challenge description \cite{diener2022interspeech}.
\subsection{Training Setup}
The models are trained over 200 epochs with a batch size of 16 and a learning rate of 5e-4. The learning rate is multiplied by 0.8 when the loss on the test set does not decrease for three consecutive epochs. The Adam optimizer is used in combination with a gradient norm clipping of 3.
The context buffer size was set to a value of 6 as a trade-off between complexity and performance. 
During training, the four most recent frames of the context buffer are sampled from the degraded signal, as well as the last two frames from the clean signal. We found that this strategy improves performance compared to  training on the degraded audio only. The training for the medium-size model takes around 6\,h on an Nvidia RTX A5000.
\subsection{Objective and Challenge Evaluation}
The first objective metric used is the wide band perceptual evaluation of speech quality (PESQ) \cite{pesq}, which was originally built to evaluate the quality of transmitted speech. The second one is DNS-MOS P.835 \cite{reddy2022dnsmos} which is a single-ended/blind measure based on a deep-learning approach for estimating Mean Opinions Scores (MOS) of speech quality (nSIG), background noise quality (nBAK) and an overall quality (nOVR). DNS-MOS was trained with ratings from previous deep-noise suppression challenges organized by Microsoft \cite{dns_2021_interspeech}. The last measure PLC-MOS \cite{diener2022interspeech} is provided by the challenge organizers. PLC-MOS utilizes a similar approach as DNS-MOS but can be used as double-ended or single-ended measure. It predicts MOS scores indicating the quality of the PLC system.

The challenge organizers evaluate the improved audio created from blind test with four metrics: Double-ended PLC-MOS, second, the DNS-MOS P.808 \cite{reddy2021dnsmos}, which returns an overall quality rating. Third, a crowd-sourced Comparison Category Rating (CCR MOS) is used, for which the participants are asked to rate the enhanced audio compared to the clean reference. This evaluation is run on the Amazon Mechanical Turk. Fourth, the word accuracy calculated by the Microsoft Azure cognitive services speech recognition is used. The final score is calculated from CCR MOS and word accuracy.
%
%
\section{Experiments and Results}
\begin{figure}[th]
  \centering
  \includegraphics[trim={4cm 12.3cm 4cm 12cm}, clip, width=1.0\linewidth]{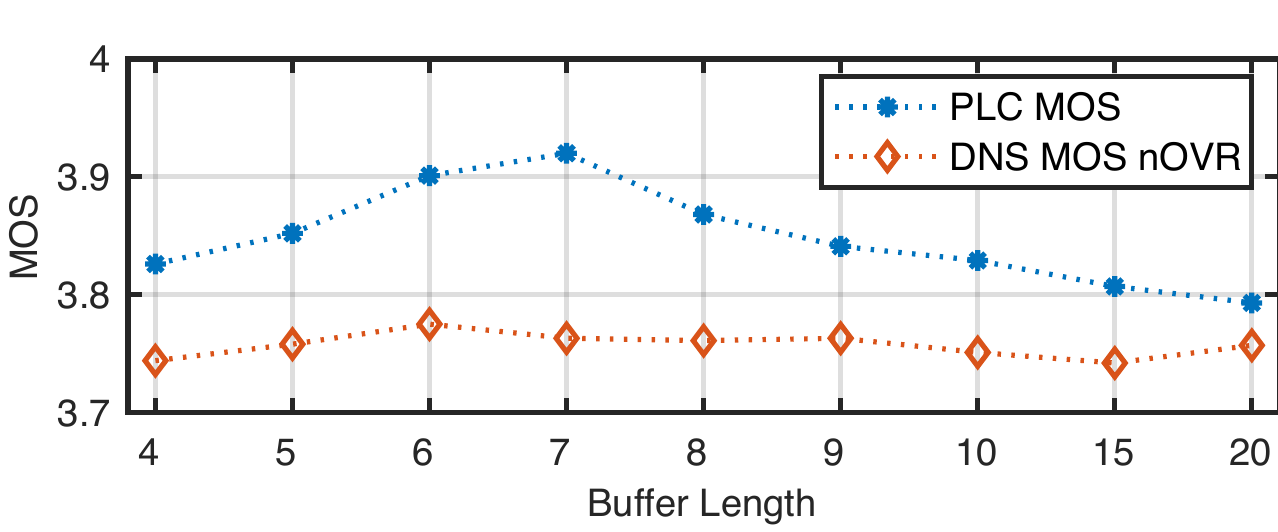}
  \caption{Performance in terms of PLC-MOS and OVR DNS-MOS P.835 versus the number of 10\,ms frames in the context buffer length.}
\label{fig:buffer_len}
\end{figure}

In the following the experiments and results are described: \\
\textbf{Objective results on complexity comparison}:
\autoref{tab:results} shows the objective results for the zero filling baseline, the feed-forward baseline trained on the same setup, as well as three complexity configurations of the tPLCnet (cf. \autoref{tab:units}). 
The FF baseline already reaches a good performance increase for all metrics. 
Performance increases with model size for all metrics.
The total improvement in terms of PLC-MOS for $\mathrm{tPLCnet_{Large}}$ is 1.08. The PESQ improvement over the zero filling baseline is 0.58 and for DNS-MOS nOVR 0.82. The PESQ and PLC-MOS improvements  are constant over all subsets.
\\
\textbf{Objective results on loss functions and loss function configurations }: We compared  $\mathrm{MAE_time}$, $\mathrm{MAE_mag}$ and $\mathrm{MAE_comb}$; for the latter, three different window sizes with 50\,\% overlap were tested to identify an optimal configuration for the PLC task. All results were acquired with $\mathrm{tPLCnet_{Medium}}$. The results are shown in \autoref{tab:loss_results}. The lowest performance in all metrics is achieved with the time-domain $\mathrm{MAE_time}$ loss. The highest PESQ  (2.76) and DNS-MOS (3.85) values are reached with $\mathrm{MAE_comb}$ loss with 20\,ms window size. The highest PLC loss (3.90) is achieved with  $\mathrm{MAE_comb}$ and 32\,ms window size.
\begin{table}[th]
  \caption{Objective scores for different loss functions. For DNS-MOS, only nOVR is shown.}
  \label{tab:loss_results}
  \centering
  \begin{tabular}{l r r r }
    \toprule
    &  & \textbf{PLC-} &  \textbf{DNS-} \\
   \textbf{loss} & \textbf{PESQ} & \textbf{MOS} & \textbf{MOS}\\
    \midrule
    $\mathrm{MAE_{time}}$ & 2.43    &     3.30  &  3.54   \\
    $\mathrm{MAE_{mag}}$ (32 ms, 50\%) & 2.75    &   3.88   & 3.79\\
    $\mathrm{MAE_{comb}}$ (64 ms, 50\%) & 2.61    &     3.53  &  3.66   \\
    $\mathrm{MAE_{comb}}$ (32 ms, 50\%) & 2.73    &     \textbf{3.90}  &  3.78   \\
    $\mathrm{MAE_{comb}}$ (20 ms, 50\%) & \textbf{2.76}    &     3.71  &  \textbf{3.85}   \\
    
    \bottomrule
  \end{tabular}
\end{table}
\\ 
\textbf{Comparison of context buffer lengths}:
The length of the context buffer is a major factor for complexity in terms of operations for our seq2one model, since the whole buffer must be processed each time by the convolutional and the BGRU layers. 
Hence, we explore the effect of different buffer lengths on performance. \autoref{fig:buffer_len} shows the performance in terms of PLC-MOS and DNS-MOS against the buffer length. 
The optimal lengths for PLC-MOS and DNS-MOS are 7 and 6, respectively, while the lowest performance is reached with 20 frames (PLC-MOS) or 15 frames (DNS-MOS). 
\\
\textbf{Results of the challenge evaluations}:
The challenge results for the zero-filling baseline and all approaches submitted to the challenge are shown in \autoref{tab:ch_results}. 
$\mathrm{tPLCnet_{Large}}$ with $\mathrm{MAE_{comb}}$ using a loss window length of 32\,ms ranked 3rd in the challenge. It improves on all metrics except for word accuracy. The 1st and 2nd ranked entries improve on all metrics (including word accuracy). Our tPLCnet achieves the second-best score in terms of PLC-MOS. 
\begin{table}[th]
  \caption{2022 INTERSPEECH Audio Deep PLC Challenge Results. Teams marked as "tied": Difference not significant ($p \geq 0.05$).}
  \label{tab:ch_results}
  \centering
  \begin{tabular}{@{}l r r r r r@{}}
    \toprule
    & \textbf{PLC-} & \textbf{DNS-} &  \textbf{CCR-}  & \textbf{Word} & \textbf{Final}\\
   \textbf{Rank} & \textbf{MOS} & \textbf{MOS} & \textbf{MOS} & \textbf{acc.} & \textbf{score}\\
    \midrule
    7 & 2.896    &     3.474  &  -1.309 & 0.860 & 0.712  \\
    Baseline & 2.904    &     3.444  &  -1.231 & 0.861 & 0.725  \\
    6 & 3.478    &     3.738  &  -1.042 & 0.825 & 0.739  \\
    5 & 3.277    &     3.513  &  -1.102 & 0.863 & 0.748  \\
    \textbf{3} (tied) \textbf{(ours)} & \textbf{3.976}    &     \textbf{3.688}  &  \textbf{-0.838} & \textbf{0.859} & \textbf{0.790}  \\
    3 (tied) & 3.829    &     3.684  &  -0.812 & 0.868 & 0.798  \\
    2nd & 3.744    &     3.788  &  -0.638 & 0.882 & 0.835  \\
    1st & 4.282    &     3.797  &  -0.552 & 0.875 & 0.845  \\
    
    \bottomrule
  \end{tabular}
\end{table}
\section{Discussion}
In the following some aspects of the results are discussed. While an increase in complexity of tPLCnet also results in a performance increase, the gain seems to saturate compared to the negative aspects of increased complexity. 
The FF baseline already produces a robust PLC performance, indicating that a large portion of the performance comes from our general approach in combination with the training setup. 
This implies that for real on-device applications the PLC model can be chosen reasonable depending on the available resources and it must not be the largest possible model. 
In \cite{cruse} a similar saturation was found for NSNET2 architecture for SE, i.e., a larger amount of RNN units heavily increases the complexity, but only slightly increases the noise reduction performance.

Regarding the comparison of loss functions, the spectral MAE emerged as the best choice, with the time-domain MAE achieving clearly lower scores. 
The magnitude-only loss reaches comparable performance to the combined version which is in line with observations in \cite{loss_functions_1}. 
Interestingly, the highest PLC-MOS is observed for a window size of 32\,ms, while the best choice for other metrics was a value of  20\,ms. 
Also,
the higher spectral resolution achieved with the 64\,ms window seems not to be helpful for PLC.

The evaluation of the length of the context buffer shows that a short context buffer with only 4 frames already reaches a good PLC performance while very long context buffers even seem detrimental. 
A similar effect for longer input buffers was observed in \cite{dnn_packet_loss}. This suggests that the main information required for robust PLC is temporally localized around the lost frame and long-term dependencies only weakly influence overall performance. However, the observed performance differences between the context buffer lengths are not very large, which also means a short buffer length can be chosen without sacrificing performance.

One observation from the challenge result is that a high PLC-MOS does not necessarily result in a high CCR-MOS. Our model reached a higher PLC-MOS than the 2nd place, but the the 2nd ranked model reached a significantly better result in the crowd-sourced CCR-MOS. We assume this can be caused by addtional speech enhancement used by the 2nd-ranked model, which modifies the spectrum such that PLC-MOS is decreased but perceived audio quality is increased. 
Word accuracies for the baseline and the tPLCnet are very close to each other (0.861 and 0.859, respectively), i.e., the PLC improvement perceived by the raters does not help the speech recognition system. We assume this is due to a slight deterioration of clean speech in the vicinity of the edges of lost frames, since our algorithm smooths the transition between these areas. 
It also seems that the speech recognition system used by challenge organizers can already handle missing packets relatively well.

\section{Conclusions}
This study introduced tPLCnet, a straightforward real-time approach for predicting lost packets from a short context buffer in the time-domain. It achieves state-of-the-art performance and ranks 3rd in the deep PLC Challenge. The training was performed with open-source data which resulted in a robust and reproducible approach which can be easily applied in the real-world.

\section{Acknowledgements}

This work was funded by the Deutsche Forschungsgemeinschaft (DFG, German Research Foundation) under Germany's Excellence Strategy – EXC 2177/1 - Project ID 390895286 and by - Project-ID 352015383 - SFB 1330.

\bibliographystyle{IEEEtran}

\bibliography{mybib}

\begin{thebibliography}{10}
\providecommand{\url}[1]{#1}
\csname url@samestyle\endcsname
\providecommand{\newblock}{\relax}
\providecommand{\bibinfo}[2]{#2}
\providecommand{\BIBentrySTDinterwordspacing}{\spaceskip=0pt\relax}
\providecommand{\BIBentryALTinterwordstretchfactor}{4}
\providecommand{\BIBentryALTinterwordspacing}{\spaceskip=\fontdimen2\font plus
\BIBentryALTinterwordstretchfactor\fontdimen3\font minus
  \fontdimen4\font\relax}
\providecommand{\BIBforeignlanguage}[2]{{%
\expandafter\ifx\csname l@#1\endcsname\relax
\typeout{** WARNING: IEEEtran.bst: No hyphenation pattern has been}%
\typeout{** loaded for the language `#1'. Using the pattern for}%
\typeout{** the default language instead.}%
\else
\language=\csname l@#1\endcsname
\fi
#2}}
\providecommand{\BIBdecl}{\relax}
\BIBdecl

\bibitem{perceptnet_dns}
J.-M. Valin, U.~Isik, N.~Phansalkar, R.~Giri, K.~Helwani, and A.~Krishnaswamy,
  ``{A Perceptually-Motivated Approach for Low-Complexity, Real-Time
  Enhancement of Fullband Speech},'' in \emph{Proc. Interspeech 2020}, 2020,
  pp. 2482--2486.

\bibitem{dtln_dns}
N.~L. Westhausen and B.~T. Meyer, ``{Dual-Signal Transformation LSTM Network
  for Real-Time Noise Suppression},'' in \emph{Proc. Interspeech 2020}, 2020,
  pp. 2477--2481.

\bibitem{perceptnet_aec}
J.-M. Valin, S.~Tenneti, K.~Helwani, U.~Isik, and A.~Krishnaswamy,
  ``Low-complexity, real-time joint neural echo control and speech enhancement
  based on percepnet,'' in \emph{ICASSP 2021 - 2021 IEEE International
  Conference on Acoustics, Speech and Signal Processing (ICASSP)}, 2021, pp.
  7133--7137.

\bibitem{dtln_aec}
N.~L. Westhausen and B.~T. Meyer, ``Acoustic echo cancellation with the
  dual-signal transformation lstm network,'' in \emph{ICASSP 2021 - 2021 IEEE
  International Conference on Acoustics, Speech and Signal Processing
  (ICASSP)}, 2021, pp. 7138--7142.

\bibitem{aec_braunschweig}
M.~M. Halimeh, T.~Haubner, A.~Briegleb, A.~Schmidt, and W.~Kellermann,
  ``Combining adaptive filtering and complex-valued deep postfiltering for
  acoustic echo cancellation,'' in \emph{ICASSP 2021 - 2021 IEEE International
  Conference on Acoustics, Speech and Signal Processing (ICASSP)}, 2021, pp.
  121--125.

\bibitem{dnn_packet_loss}
B.-K. Lee and J.-H. Chang, ``Packet loss concealment based on deep neural
  networks for digital speech transmission,'' \emph{IEEE/ACM Transactions on
  Audio, Speech, and Language Processing}, vol.~24, no.~2, pp. 378--387, 2016.

\bibitem{nsnet2}
S.~Braun and I.~Tashev, ``Data augmentation and loss normalization for deep
  noise suppression,'' in \emph{Speech and Computer}, A.~Karpov and
  R.~Potapova, Eds.\hskip 1em plus 0.5em minus 0.4em\relax Cham: Springer
  International Publishing, 2020, pp. 79--86.

\bibitem{cruse}
S.~Braun, H.~Gamper, C.~K. Reddy, and I.~Tashev, ``Towards efficient models for
  real-time deep noise suppression,'' in \emph{ICASSP 2021 - 2021 IEEE
  International Conference on Acoustics, Speech and Signal Processing
  (ICASSP)}, 2021, pp. 656--660.

\bibitem{mohamed2020concealnet}
M.~M. Mohamed and B.~W. Schuller, ``Concealnet: An end-to-end neural network
  for packet loss concealment in deep speech emotion recognition,'' \emph{arXiv
  preprint arXiv:2005.07777}, 2020.

\bibitem{plc_lstm}
R.~Lotfidereshgi and P.~Gournay, ``Speech prediction using an adaptive
  recurrent neural network with application to packet loss concealment,'' in
  \emph{2018 IEEE International Conference on Acoustics, Speech and Signal
  Processing (ICASSP)}, 2018, pp. 5394--5398.

\bibitem{facebook_plc}
J.~Lin, Y.~Wang, K.~Kalgaonkar, G.~Keren, D.~Zhang, and C.~Fuegen, ``A
  time-domain convolutional recurrent network for packet loss concealment,'' in
  \emph{ICASSP 2021 - 2021 IEEE International Conference on Acoustics, Speech
  and Signal Processing (ICASSP)}, 2021, pp. 7148--7152.

\bibitem{plc_gan_1}
\BIBentryALTinterwordspacing
J.~Wang, Y.~Guan, C.~Zheng, R.~Peng, and X.~Li, ``A temporal-spectral
  generative adversarial network based end-to-end packet loss concealment for
  wideband speech transmission,'' \emph{The Journal of the Acoustical Society
  of America}, vol. 150, no.~4, pp. 2577--2588, 2021. [Online]. Available:
  \url{https://doi.org/10.1121/10.0006528}
\BIBentrySTDinterwordspacing

\bibitem{inpainting_gan}
H.~Zhou, Z.~Liu, X.~Xu, P.~Luo, and X.~Wang, ``Vision-infused deep audio
  inpainting,'' in \emph{Proceedings of the IEEE/CVF International Conference
  on Computer Vision (ICCV)}, October 2019.

\bibitem{plc_gan_2}
Y.~Shi, N.~Zheng, Y.~Kang, and W.~Rong, ``Speech loss compensation by
  generative adversarial networks,'' in \emph{2019 Asia-Pacific Signal and
  Information Processing Association Annual Summit and Conference (APSIPA
  ASC)}, 2019, pp. 347--351.

\bibitem{diener2022interspeech}
L.~Diener, S.~Sootla, S.~Branets, A.~Saabas, R.~Aichner, and R.~Cutler,
  ``Interspeech 2022 audio deep packet loss concealment challenge,'' in
  \emph{{INTERSPEECH} 2022 - 23rd Annual Conference of the International Speech
  Communication Association}, 2022 (submitted).

\bibitem{dns_challenge}
C.~K. Reddy, V.~Gopal, R.~Cutler, E.~Beyrami, R.~Cheng, H.~Dubey,
  S.~Matusevych, R.~Aichner, A.~Aazami, S.~Braun, P.~Rana, S.~Srinivasan, and
  J.~Gehrke, ``{The INTERSPEECH 2020 Deep Noise Suppression Challenge:
  Datasets, Subjective Testing Framework, and Challenge Results},'' in
  \emph{Proc. Interspeech 2020}, 2020, pp. 2492--2496.

\bibitem{AEC_challenge_ver1}
K.~Sridhar, R.~Cutler, A.~Saabas, T.~Parnamaa, M.~Loide, H.~Gamper, S.~Braun,
  R.~Aichner, and S.~Srinivasan, ``Icassp 2021 acoustic echo cancellation
  challenge: Datasets, testing framework, and results,'' in \emph{ICASSP 2021 -
  2021 IEEE International Conference on Acoustics, Speech and Signal Processing
  (ICASSP)}, 2021, pp. 151--155.

\bibitem{dns_2021_interspeech}
Z.~Xu, M.~Strake, and T.~Fingscheidt, ``{Deep Noise Suppression with
  Non-Intrusive PESQNet Supervision Enabling the Use of Real Training Data},''
  in \emph{Proc. Interspeech 2021}, 2021, pp. 2806--2810.

\bibitem{tasnet_orig}
Y.~Luo and N.~Mesgarani, ``Tasnet: Time-domain audio separation network for
  real-time, single-channel speech separation,'' in \emph{2018 IEEE
  International Conference on Acoustics, Speech and Signal Processing
  (ICASSP)}, 2018, pp. 696--700.

\bibitem{conv_tasnet}
------, ``Conv-tasnet: Surpassing ideal time–frequency magnitude masking for
  speech separation,'' \emph{IEEE/ACM Transactions on Audio, Speech, and
  Language Processing}, vol.~27, no.~8, pp. 1256--1266, 2019.

\bibitem{dua-path}
C.~Li, Y.~Luo, C.~Han, J.~Li, T.~Yoshioka, T.~Zhou, M.~Delcroix, K.~Kinoshita,
  C.~Boeddeker, Y.~Qian, S.~Watanabe, and Z.~Chen, ``Dual-path rnn for long
  recording speech separation,'' in \emph{2021 IEEE Spoken Language Technology
  Workshop (SLT)}, 2021, pp. 865--872.

\bibitem{chung2014empirical}
J.~Chung, C.~Gulcehre, K.~Cho, and Y.~Bengio, ``Empirical evaluation of gated
  recurrent neural networks on sequence modeling,'' \emph{arXiv preprint
  arXiv:1412.3555}, 2014.

\bibitem{loss_functions_1}
S.~Braun and I.~Tashev, ``A consolidated view of loss functions for supervised
  deep learning-based speech enhancement,'' in \emph{2021 44th International
  Conference on Telecommunications and Signal Processing (TSP)}, 2021, pp.
  72--76.

\bibitem{loss_functions_2}
S.~Braun and H.~Gamper, ``Effect of noise suppression losses on speech
  distortion and asr performance,'' \emph{arXiv preprint arXiv:2111.11606},
  2021.

\bibitem{wang17n_interspeech}
Y.~Wang, R.~Skerry-Ryan, D.~Stanton, Y.~Wu, R.~J. Weiss, N.~Jaitly, Z.~Yang,
  Y.~Xiao, Z.~Chen, S.~Bengio, Q.~Le, Y.~Agiomyrgiannakis, R.~Clark, and R.~A.
  Saurous, ``{Tacotron: Towards End-to-End Speech Synthesis},'' in \emph{Proc.
  Interspeech 2017}, 2017, pp. 4006--4010.

\bibitem{psa_1}
B.~Patton, J.~Skoglund, J.~Thorpe, J.~Hershey, K.~Wilson, M.~Chinen, R.~F.
  Lyon, and R.~A. Saurous, ``Exploring tradeoffs in models for low-latency
  speech enhancement,'' in \emph{Proceedings of the 16th International Workshop
  on Acoustic Signal Enhancement}, 2018.

\bibitem{pesq}
``{ITU-T P.862: Perceptual evaluation of speech quality (PESQ): An objective
  method for end-to-end speech quality assessment of narrow-band telephone
  networks and speech codecs.}'' 2001.

\bibitem{reddy2022dnsmos}
C.~K. Reddy, V.~Gopal, and R.~Cutler, ``Dnsmos p.835: A non-intrusive
  perceptual objective speech quality metric to evaluate noise suppressors,''
  in \emph{ICASSP}, 2022.

\bibitem{reddy2021dnsmos}
C.~K.~A. Reddy, V.~Gopal, and R.~Cutler, ``Dnsmos: A non-intrusive perceptual
  objective speech quality metric to evaluate noise suppressors,'' in
  \emph{ICASSP 2021 - 2021 IEEE International Conference on Acoustics, Speech
  and Signal Processing (ICASSP)}, 2021, pp. 6493--6497.

\end{thebibliography}

\end{document}